\documentclass[prb,showpacs,superscriptaddress,floatfix]{revtex4}

\usepackage{graphicx}
\usepackage{amsmath,amstext,bm}
\usepackage{amsfonts}

\renewcommand{\vec}{\bm}

\begin{document}
\title{Ultrafast Spin Dynamics in Optically Excited Bulk GaAs at Low Temperatures}


\author{M. Krau\ss}
\affiliation{Physics Department and
  Research Center OPTIMAS, University of Kaiserslautern, P. O. Box
  3049, 67663 Kaiserslautern, Germany} 
\author{R. Bratschitsch}
\altaffiliation{Present address: University of Konstanz and Center for
  Applied Photonics, 78457 Konstanz, Germany}
\affiliation{JILA, National Institute of Standards and Technology and
  University of Colorado, Boulder, CO 80390-0440, USA}
\author{Z. Chen}
\affiliation{JILA, National Institute of Standards and Technology and
  University of Colorado, Boulder, CO 80390-0440, USA}
\author{S. T. Cundiff}
\affiliation{JILA, National Institute of Standards and Technology and
  University of Colorado, Boulder, CO 80390-0440, USA}
\author{H. C. Schneider}
\homepage{http:\\www.physik.uni-kl.de\schneider}
\affiliation{Physics Department and
  Research Center OPTIMAS, University of Kaiserslautern, P. O. Box
  3049, 67663 Kaiserslautern, Germany}

\date{\today}


\begin{abstract}

This paper presents a study of electron spin dynamics in \emph{bulk}
GaAs at low temperatures for elevated optical excitation
conditions. Our time-resolved Faraday rotation measurements yield
sub-nanosecond electron spin dephasing times over a wide range of
n-doping concentrations in quantitative agreement with a microscopic
treatment of electron spin dynamics. The calculation shows the
occurrence and breakdown of motional narrowing for spin dephasing
under elevated excitation conditions. We also find a peak of the spin
dephasing time around a doping density for which, under lower
excitation conditions, a metal-insulator transition occurs. However,
the experimental results for high excitation can be explained without
a metal-insulator transition. We therefore attribute the peak in
spin-dephasing times to the influence of screening and scattering on
the spin-dynamics of the excited electrons.

\end{abstract}




\maketitle

\section{Introduction}

Electron spin dynamics in semiconductors has experienced a dramatic
revival of interest in the last decade due to the promise of
semiconductor-based spintronics
applications.~\cite{awschalom-book,zutic:review} In contrast to
magneto-electronic systems involving metallic multilayers,
semiconductors are better suited for the fabrication of integrated
electronic and opto-electronic devices, but in non-ferromagnetic
semiconductors the spin is not a conserved quantity due to the
spin-orbit interaction. Thus an understanding of spin relaxation in
semiconductors is essential, because it limits all information
processing and storage using electronic spins. The renewed interest
in combination with improved experimental techniques has led to
important new results. For instance, long spin dephasing times in n-doped
GaAs at low temperatures and low carrier densities have been measured
using time-resolved Faraday rotation (FR)
techniques.~\cite{kikkawa:prl98,heberle:prl01} Very recently, the
technique of spin noise measurements has provided a new tool for the
determination of ``intrinsic'' spin relaxation times without
electrical or optical carrier
injection.~\cite{oestreich-spinnoise-qw:prl08} These measurements
show, for instance, the importance of a metal-insulator transition on
the spin relaxation at low temperatures. However, the foundation for
the theoretical explanation of these experiments above the transition
temperature was already laid by an analysis of Dyakonov and Perel
(DP)~\cite{dyakonov-perel:jetp71} who showed that in n-doped GaAs the
decay of spin polarization occurs through the dephasing of the spin
coherence in a spatially inhomogeneous effective magnetic field
provided by the band structure.

Although these ``non-invasive'' experimental techniques yield
important results,~\cite{dzhioev-spin-relaxation:prb02} one may also
be interested in the excitation of appreciable spin-polarized carrier
densities by optical or electronic means. Existing studies, which
address these questions, have concentrated on the electron spin
dynamics in n-doped GaAs-based semiconductors under electrical bias
and Hanle effect measurements on a timescale of several hundred
picoseconds.~\cite{crooker:apl06-spin-relaxation-times-with-bias}

This paper is concerned with an investigation of the spin-dephasing
dynamics in n-doped bulk GaAs after the excitation of spin-polarized
carrier densities by ultrashort optical pulses. For a wide range of
doping densities, we present a detailed comparison between
time-resolved FR measurements and a microscopic theory, which is based
on the original DP analysis. Our approach to compute the
spin-dephasing time does not require the determination of auxiliary
quantities, such as an effective momentum scattering time, which is
often employed to achieve agreement with
experiment.~\cite{beck-epl06:spin-lifetime} Instead, the equations of
motion for the spin and momentum-resolved electronic distribution
functions are solved numerically including the relevant effects of the
band-structure and the scattering of electrons with other electrons,
phonons, and impurities.~\cite{wu-euphysb00:dp-qw} Due to the
numerical complexity of calculating the spin dynamics in \emph{bulk}
GaAs, theoretical results so far have concentrated on the microscopic
spin dynamics in low-dimensional
semiconductors,~\cite{lau:spin-coherence:prb01,schuller-spin-relaxation-qw:prl07}
with the exception of recent work,~\cite{jiang-wu:prb09} or on
accurate calculations of the analytical DP results for the spin
decay.~\cite{ridley-lophon,schilf-prb05}

This paper is organized as follows. In  Sec.~\ref{sec:theory} we
explain the theoretical approach and present the dynamical equations
for the spin-density matrix. In Sec.~\ref{sec:experiment} we describe
the experimental setup and our modeling of the experimental conditions
in the framework of the theory. Sec.~\ref{sec:results} contains the
experimental results and their interpretation by comparing them with
model calculations.

\section{Theoretical Approach}
\label{sec:theory}

We directly calculate the dynamics of the momentum resolved $2\times2$
spin-density matrix 
\begin{equation}
\rho_{ss'}(\vec{k})=\left\langle c_{s,\vec k}^\dagger c_{s'\vec k} \right\rangle
\end{equation}
where $s=\,\uparrow,\downarrow$ denotes the spin projection quantum
number along the quantization axis~$z$, $c$ ($c^\dag$) are electron
destruction (creation) operators in the Heisenberg picture, and
$\langle\cdots\rangle$ denotes the average with respect to the
equilibrium statistical operator. The elements of the spin-density
matrix are the carrier occupation numbers $n^{s}_{\vec
  k}=\rho_{ss}(\vec k)$ and the complex coherences $\Psi_{\vec
  k}=\rho_{\uparrow\downarrow}(\vec k)$.  In terms of the spin
density-matrix, the average spin is given by
\begin{align}
\langle s_x(\vec k)
\rangle = & \frac{\hbar}{2} \mathrm{Re} [\Psi_{\vec{k}}] \\
\langle s_y(\vec k)
\rangle = &\frac{\hbar}{2} \mathrm{Im}[\Psi_{\vec{k}}] \\
\langle s_z(\vec k) \rangle = &\frac{\hbar}{2} (n^{\uparrow}_{\vec{k}}-n^{\downarrow}_{\vec{k}}) .
\end{align}
The time evolution of $\rho(\vec{k})$ is determined by the
equation of motion
\begin{equation}
\frac{\partial \rho(\vec{k})}{\partial t} = 
\frac{\partial \rho(\vec{k})}{\partial t}\Bigr|_{\text{coh}} + 
\frac{\partial \rho(\vec{k})}{\partial t}\Bigr|_{\text{scatt}}
\label{HEM}
\end{equation}
where the coherent part includes the electronic energies, the
influence of the effective magnetic field due to the band structure,
and Hartree-Fock renormalizations. For the diagonal elements
$n_{\vec k}^s$ we have 
\begin{align}
\label{n-coherent}
\frac{\partial}{\partial t} n_{\vec k}^s \bigr|_\mathrm{coh} =\mbox{}& -\frac{2s}{\hbar}\{[g\mu_B B 
					+ \Omega_x(\vec{k})] \mathrm{Im}(\Psi_{\vec{k}}) 
					+\Omega_y(\vec{k}) \mathrm{Re}(\Psi_{\vec{k}})	\}  \nonumber \\
				  &+ \frac{4s}{\hbar} \mathrm{Im} \sum\limits_{\vec{q}} W_{\vec q} \Psi_{\vec{k} + \vec{q}}^* \Psi_{\vec{k}} 
\end{align}
andthe off-diagonal elements are given by
\begin{align}
\label{Psi-coherent}
\frac{\partial}{\partial t} \Psi_{\vec k} \bigr|_\mathrm{coh} = \mbox{}&
		\frac{1}{2\hbar} [ ig\mu_B B+i\Omega_x(\vec k) + \Omega_y(\vec k) ] (n_{\vec k}^{\uparrow} - n_{\vec k}^{\downarrow} ) - \frac{i}{\hbar} \Omega_z(\vec k) \Psi_{\vec k} \nonumber \\
	&+ \frac{i}{\hbar} \sum\limits_{\vec q} W_{\vec q} [ 
			(n_{\vec k+\vec q}^{\uparrow} - n_{\vec k+\vec q}^{\downarrow} ) \Psi_{\vec k} 
			 - \Psi_{\vec k+\vec q}(n_{\vec k}^{\uparrow} - n_{\vec k}^{\downarrow} )] .
\end{align}
Here, $s=\pm1/2$ denotes the electron spin and $\vec B = B \hat e_x$ in the Voigt geometry.
The screened Coulomb interaction 
\begin{equation}
W_{\vec q} =
\frac{1}{\tilde\varepsilon_{\mathrm{bg}}\tilde\varepsilon_{\vec{q}}} \,v_q
\end{equation}
in Eqs.~\eqref{n-coherent} and~\eqref{Psi-coherent} contains the bare
Coulomb potential
\begin{equation}
v_{\vec{q}}=\frac{e^2}{\varepsilon_0} \frac{1}{q^2}
\end{equation}
and the dimensionless dielectric functions
$\tilde\varepsilon_{\mathrm{bg}}$ and $\tilde\varepsilon_{\vec{q}}$. The
former describes the ``background'' screening due to the
polarizability of the lattice, and the latter the screening due to the
electrons provided by the doping. For this, we use the static limit of
the Lindhard dielectric function~\cite{haug-koch} in the numerical
evaluation.  The scattering contributions (electron-electron,
electron-phonon, and electron-impurity) to Eq.~\eqref{HEM} can be
obtained by equation-of-motion or Green's function
techniques~\cite{wu:prb00:kinetics_qw,glazov-jetp04,lechner-prb05:spin-relaxation-times}
following, e.g.,
Refs.~\onlinecite{schaefer-book,binder-koch,rossi:rmp02}.  As an
example we write out the expressions for Coulomb scattering and define
the kernels for in- and out-scattering
\begin{equation}
K^{\mathrm{in}}_{\vec k_1,\vec q} = (1-n^{\uparrow}_{\vec k_1 + \vec q})n^{\uparrow}_{\vec k_1} + (1-n^{\downarrow}_{\vec k_1 + \vec q})n^{\downarrow}_{\vec k_1}
- 2\mathrm{Re}(\Psi_{\vec k_1 + \vec q}^{*} \Psi_{\vec k_1})
\end{equation}
and
\begin{equation}
K^{\mathrm{out}}_{\vec k_1,\vec q} = n^{\uparrow}_{\vec k_1 + \vec q}(1-n^{\uparrow}_{\vec k_1}) + n^{\downarrow}_{\vec k_1 + \vec q}(1-n^{\downarrow}_{\vec k_1}) 
- 2\mathrm{Re}(\Psi_{\vec k_1 + \vec q}^{*} \Psi_{\vec k_1})
\end{equation}
so that the scattering contribution for the diagonal elements becomes
\begin{align}
\label{dn-dt}
\frac{\partial}{\partial t} n_{\vec k}^s \bigr|_\mathrm{scat} 
= \mbox{}&\frac{2\pi}{\hbar} \sum\limits_{\vec k_1,\vec q} W_{\vec q}^2 
\delta(\epsilon_{\vec k}-\epsilon _{|\vec{k}+\vec{q}|}+\epsilon_{|\vec{k}_{1}+\vec{q}|}-\epsilon_{\vec k_{1}})  \nonumber \\
&\times\bigl(K^{\mathrm{in}}_{\vec k_1,\vec q}[(1-n_{\vec k}^s)n_{\vec k+\vec q}^s-\mathrm{Re}(\Psi_{\vec k}^* \Psi_{\vec k + \vec q})] \\
& - K^{\mathrm{out}}_{\vec k_1,\vec q}[n_{\vec k}^s(1-n_{\vec k+\vec q}^s)-\mathrm{Re}(\Psi_{\vec k}^* \Psi_{\vec k + \vec q})]\bigr) . \nonumber
\end{align}
For the off-diagonal elements we have
\begin{align}
\label{dPsi-dt}
\frac{\partial}{\partial t}  \Psi_{\vec k} \bigr|_\mathrm{scat} = \mbox{}&\frac{\pi}{\hbar} \sum\limits_{\vec k_1,\vec q} W_{\vec q}^2 \delta(\epsilon _{\vec k}-\epsilon _{|\vec{k}+\vec{q}|}+\epsilon_{|\vec{k}_{1}+\vec{q}|}-\epsilon_{\vec k_{1}})  \nonumber \\
&\times\bigl(K^{\mathrm{in}}_{\vec k_1,\vec q}[\Psi_{\vec k + \vec q}(2-n_{\vec k}^{\uparrow}+n_{\vec k}^{\downarrow})-\Psi_{\vec k}(n_{\vec k+\vec q}^{\uparrow}+n_{\vec k+\vec q}^{\downarrow})]  \\
&- K^{\mathrm{out}}_{\vec k_1,\vec q}[\Psi_{\vec k}(2-n_{\vec k + \vec q}^{\uparrow}+n_{\vec k + \vec q}^{\downarrow})-\Psi_{\vec k + \vec q}(n_{\vec k}^{\uparrow}+n_{\vec k}^{\downarrow})]\bigr) \nonumber.
\end{align}
In Eqs.~\eqref{dn-dt} and~\eqref{dPsi-dt}, the electronic energy
dispersions are given by $\epsilon_k \equiv \epsilon^{\mathrm{e}}_k =
\hbar^2/(2m)\,k^2$ with the electron effective mass $m_{\mathrm{e}}=
0.067m_0$. An important step for the numerical solution of
Eq.~\eqref{HEM} for \emph{bulk} GaAs is to account for the strongly
anisotropic intrinsic magnetic field $\vec{\Omega}(\vec k)$ by an
expansion of $\rho(\vec{k})$ in a set of orthonormal functions
$L_{\ell}(\vartheta,\varphi)$. Such an expansion was already used in
Ref.~\onlinecite{pikus-titkov:opt-orient} but, in contrast to the
original treatment, we choose an expansion in functions that describe
the angular dependence of $\vec{\Omega}$ most accurately, namely:
$L_1(\vartheta,\varphi)\propto 1$ and
$L_{2,3,4}(\vartheta,\varphi)\propto \Omega_{x,y,z}$ where the
components of the intrinsic magnetic field are given by
\begin{align}
\label{Omega-k3}
\Omega_x(\vec k) &= \gamma k_x(k_y^2-k_z^2) \\
\Omega_y(\vec k) &= \gamma k_y(k_z^2-k_x^2) \\
\Omega_z(\vec k) &= \gamma k_z(k_x^2-k_y^2) 
\end{align}
with a material specific constant~$\gamma$.~\cite{winkler:book}

Before we turn to the numerical results from this theory, which
describes nonequilibrium spin dynamics due to the Dyakonov-Perel
mechanism, it should be noted that there exists a variety of effects
which lead to relaxation and dephasing of spin polarizations in GaAs.
However, in moderately n-doped GaAs at low temperatures under elevated
optical excitation conditions, the Dyakonov-Perel mechanism is
dominant. The quantitative agreement between the parameter free
microscopic calculation and the Faraday rotation measurements
additionally, presented in Sec.~\ref{sec:results}, supports this point
of view. For definiteness, we mention the following competing
processes: Besides the intrinsic magnetic field $\vec{\Omega}$, the
$g$-factor dispersion of electrons in GaAs~\cite{zutic:review} leads
to additional dephasing. We neglect this effect because the optically
excited spin polarization is due to electrons in a small energy
interval above the Fermi energy (less than 10\,meV). In this
region the spread of rotation frequencies due to the $g$-factor dispersion
$\Delta g \mu_\mathrm{B} |\vec B| $ is typically one order of
magnitude smaller than the internal magnetic fields $\hbar \langle |
\vec \Omega(\vec k) | \rangle$. Thus the dephasing due to the
$g$-factor dispersion is of minor importance for the theoretical
analysis of the experiments presented in this paper. The
Bir-Aronov-Pikus mechanisms usually acts on a longer timescale for the
relatively low density of optically excited holes and therefore does
not play a dominant role,
either.~\cite{maialle:prb96,hcs-prb06:spin-relax-surface} Relaxation
due to interactions with a nuclear spin polarization is not included
in the calculation; in the experimental results it is excluded by
periodically switching the helicity of the circularly polarized
excitation beam.

\begin{figure}[t!]
\includegraphics[width=0.45\textwidth]{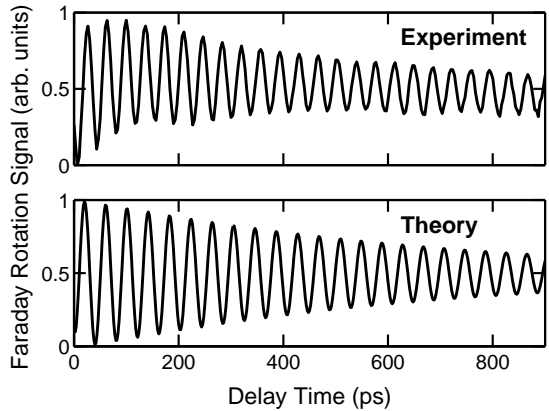}
\caption{Time-resolved Faraday rotation signal for
  $N_\mathrm{dop}=3.6\times10^{16}\,\mathrm{cm}^{-3}$, $B=4$\,T, and
  $T=4$\,K as obtained by experiment (top), and numerical calculation
  (bottom). For the calculation
  $N_\mathrm{exc}=1\times10^{16}\,\mathrm{cm}^{-3}$ was assumed. The
  spin-dephasing time $T_{2}^*$ is determined by the exponential decay
  time of the envelope.}
\end{figure}

\section{Experiment}
\label{sec:experiment}

We have performed ultrafast FR measurements to determine the electron
spin dynamics in n-doped bulk GaAs
layers.~\cite{awschalom-book,Rudi-apl06,Rudi-nat07} The doped GaAs
wafer was mounted on a sapphire substrate and thinned down to about
30\,$\mu$m for the transmission measurement. Spin polarized electrons
were optically excited by a circularly polarized pump pulse.  In this
setup, the ultrafast laser pulses are generated by a mode-locked
Ti:sapphire laser oscillator operating at 76\,MHz repetition
frequency. The samples were held at low temperatures in an optical
cryostat with a split-coil superconducting magnet, and a magnetic
field of typically 4\,T was applied. After the excitation, the
electron spins begin to precess in a magnetic field applied parallel
to the sample surface (Voigt geometry). By measuring the polarization
rotation of a linearly polarized probe beam of the same wavelength
transmitted through the sample, the electron spin dynamics can be
monitored directly. The spin precession results in an oscillating
signal, as shown in Fig.~1, that decays exponentially. We use an
exponential fit for the signal envelope to extract the decay time due
to dephasing processes, which is conventionally
denoted~\cite{zutic:review} by the ensemble spin dephasing time~
$T_2^*$.

For the comparison with the experiment, we use for the calculation
initial carrier distributions $n^0_k = n^{\mathrm{dop}}_k +
n^{\mathrm{exc}}_k$, which contain both the influence of the doping
and the optical excitation, because the excitation pulse duration is
much shorter than the timescale of the subsequent electronic spin
dynamics, so that any coherences introduced by the pulse can be
ignored for the electronic dynamics. The itinerant carriers introduced
by the doping are assumed to be completely relaxed, and are modeld by
unpolarized Fermi-Dirac electron distributions $n^{\mathrm{dop}} =
f(\epsilon^{\mathrm{e}}_k)$ at lattice temperature with electronic
density equal to the density of dopants. We model the electronic
carrier distribution $n^{\mathrm{exc}}_{k}$ excited by the ultrashort
circularly polarized laser pulse propagating in $z$~direction as a
Gaussian with spectral width of the pulse, $\sigma = 15$\,meV. In
accordance with the experimental procedure, we take the photon energy
$\hbar\omega_{\mathrm{p}}$ approximately equal to the electron-hole
transition energy at the Fermi momentum of the doping density under
consideration. As in the experiment, we include a magnetic
field of 4\,T, a lattice temperature 4\,K. Further we choose a 50\%
initial spin polarization of the optically excited carriers, which
corresponds to the optimal spin polarization achievable by optical
orientation at the bandgap.~\cite{pikus-titkov:opt-orient} For the
calculation of the time-resolved FR signal shown in Fig.~1, we assume
a linearly polarized probe pulse degenerate with the circularly
polarized pump. Assuming that the holes are
unpolarized~\cite{hilton-tang,michael-prl08} on the timescale of
interest and that excitonic effects do not contribute, the FR signal
is determined by~\cite{sham-spin-beatings:prl95}
\begin{equation}
\Theta_{\text{F}} \propto \sum_{\vec{k}} \, g(\vec{k})
[n^\uparrow_{\vec{k}}-n^\downarrow_{\vec{k}}]
\end{equation}
Here, $g(\vec{k}) \propto \exp[(\epsilon^{\mathrm{e}}_k -
\epsilon^{\mathrm{h}}_k-\hbar\omega_{\mathrm{p}})^2/\sigma^2]$, with
$\sigma$ the spectral width of the pulse and $\epsilon^{\mathrm{h}}_k
= \hbar^2k^2/(2m_{\mathrm{h}})$, $m_{\mathrm{h}}= 0.457m_0$, contains
the contribution of the electron-hole transition at wavevector $\vec
k$ to the signal at the probe photon
energy~$\hbar\omega_{\mathrm{p}}$. For the experimental and
theoretical ``raw data,'' as shown, for instance, in Fig.~1, we use the
same exponential fit for the decay of the signal envelope to
extract~$T_2^*$.

\section{Results}
\label{sec:results}

Figure~2 shows the dependence of the spin-dephasing time $T_{2}^{*}$
on the photoexcited carrier density. Both in theory and experiment we
find monotonically decreasing dephasing times for n-doped GaAs
($N_\mathrm{dop}= 3.6\times10^{16}$\,cm$^{-3}$) and increasing
dephasing times for undoped GaAs with the better agreement for the
undoped GaAs. When comparing measured and calculated results, it
should be noted that there is a considerable uncertainty in the
experiment because the photoexcited carrier density is difficult to
estimate, and for the doped sample, we have an additional uncertainty
in the doping density. Taking this into account, the agreement for the
doped GaAs is very good. The agreement for undoped GaAs is even better
because the weak dependence of the spin-dephasing time on the excited
carrier density means that the experimental uncertainty in the
optically excited carrier density has only a limited influence.

The spin-dephasing times for undoped and n-doped GaAs in Fig.~2 differ
by more than a factor of three and have different slopes due to the
different spin dynamics in both cases: In the undoped material, the
spin of a nonequilibrium, highly spin-polarized (50\%) electron plasma
relaxes, whereas in the doped material the spin dynamics of polarized
optically excited electrons occurs in the presence of unpolarized
electrons residing in the material due to the n-doping. For these two
scenarios, the interplay of the different scattering mechanisms
(carrier-carrier, carrier-phonon, and carrier-impurity) and screening,
and thus the influence of the optically excited carrier density on the
spin-dephasing time are different.

\begin{figure}[t]
\includegraphics[width=0.45\textwidth]{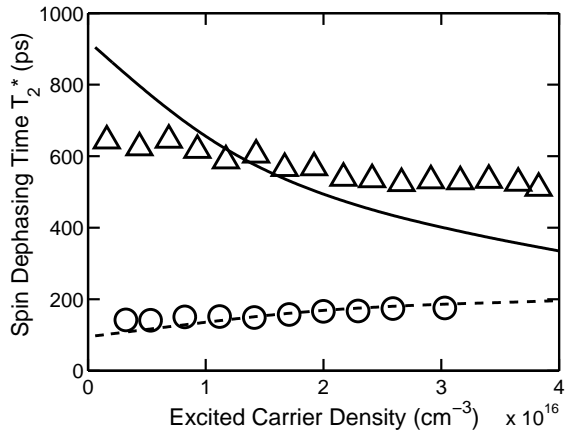}
\caption{Measured (symbols) and calculated (lines) spin dephasing
  times vs.\ photoexcited carrier density for doped (upper curves) and
  undoped (lower curves) bulk GaAs at $T =4$\,K and with $B=4$\,T
  external magnetic field. The doping density is
  $3.6\times10^{16}\,\mathrm{cm}^{-3}$ (calculation) and $(3.6\pm 0.3)
  \times10^{16}\,\mathrm{cm}^{-3}$ (experiment).}
\end{figure}

Figure~2 already shows the pronounced influence of the doping density
on the spin-dephasing time, and we investigate it now in more
detail. The theoretical and experimental results in Fig.~3 are
obtained by varying the doping density for a fixed density of
optically excited carriers of $2\times 10^{16}$\,cm$^{-3}$. We find
good agreement of the measured data points with our microscopic
calculation including only itinerant-electron dynamics, which suggests
that the peak of the spin-dephasing time at intermediate doping
densities can be explained without invoking a metal-insulator
transition.~\cite{dzhioev-spin-relaxation:prb02} A qualitative
explanation of this peak is provided by the motional narrowing
relation~\cite{zutic:review,brand-spin-precession-qw:prl02}
\begin{equation}
\tau_{\mathrm{spin}} \propto \frac{1}{\tau_{\mathrm{p}}} \quad \text{if}
  \quad \langle \Omega \rangle \tau_{\mathrm{p}} \ll 1 .
\label{motional-narrowing}
\end{equation} 
In Eq.~\eqref{motional-narrowing}, $\langle \Omega \rangle$ is the
effective Larmor frequency due to the DP spin splitting averaged over
the electrons taking part in the spin-precession and dephasing
dynamics. The spin and momentum relaxation times are denoted by
$\tau_{\mathrm{spin}}$ and $\tau_{\mathrm{p}}$, respectively.  For low
and moderate doping densities, i.e., for densities up to
$10^{17}$\,cm$^{-3}$ we are in the motional narrowing regime because
the Larmor frequency $\bm\Omega(\vec k)$ of the precession of an
electron is proportional to~$k^3$. Thus it becomes large only for
electrons excited at larger momenta, which is the case at higher
doping concentrations for optical excitation beyond the quasi
Fermi-level of the relaxed electrons. The combination of elastic
impurity scattering and mainly inelastic Coulomb scattering leads to a
decrease of the effective momentum scattering time~$\tau_{\mathrm{p}}$
with increasing doping density. Via the motional narrowing
relation~\eqref{motional-narrowing} this translates to a longer
spin-dephasing time. For higher doping concentrations, the motional
narrowing regime described by the inequality in
Eq.~\eqref{motional-narrowing} is left and the spin-dephasing time
drops with increasing doping concentration.~\cite{zutic:review}

For a more quantitative understanding of the doping density dependence
in Fig.~3, we analyze the contributions of the different scattering
mechanisms. To this end, we also show in Fig.~3 the spin-dephasing time
for the two artificial cases of (i) only elastic impurity scattering,
and (ii) no impurity scattering. For case (ii) we still include
carrier-carrier and carrier-phonon scattering, which lead to elastic
and inelastic scattering events, but the inelastic processes have the
larger available phase space and therefore dominate case (ii). For
undoped GaAs, i.e., doping densities of less than
$10^{15}$\,cm$^{-3}$, the full calculation is identical to case (ii),
which shows that carrier-carrier and carrier-phonon scattering are
responsible for the spin dephasing.~\footnote{Curve (i) is not plotted
  for densities below $4 \times 10^{15}$\,cm$^{-3}$ because for low
  impurity concentrations there are hardly any impurity scattering
  events contributing to spin dephasing, so that the spin-dephasing
  time rapidly approaches a timescale of 10\,ns.}  For intermediate
doping densities between $10^{15}$\,cm$^{-3}$ and $10^{17}$\,cm$^{-3}$
the rise of the spin-dephasing time towards a maximum at $8\times
10^{16}$\,cm$^{-3}$ is determined by the \emph{interplay} of the
elastic impurity scattering, the predominantly inelastic
carrier-carrier/phonon scattering, and the influence of screening on
all three scattering processes. Inelastic or elastic scattering
processes alone cannot explain the behavior of the spin-dephasing time
in this region: If only elastic scattering is included, as in case
(i), electrons are not scattered away from the higher $k$ states, in
which they are excited by the laser; there they experience a faster
precession, cf.~Eq.~\eqref{Omega-k3}, and thus dephase more quickly
with increasing doping concentration. This effect counteracts the
increasing effectiveness of the carrier-impurity scattering, which, in
the picture of Eq.~\eqref{motional-narrowing}, leads to a shorter
$\tau_{\mathrm{p}}$. If only carrier-carrier and carrier-phonon
scattering is included, as in case (ii), the optically excited carrier
density of $2\times 10^{16}$\,cm$^{-3}$ dominates over or is
comparable to the doped carrier density for doping concentration of up
to several $10^{16}$\,cm$^{-3}$, so that no pronounced density
dependence is to be expected for these scattering mechanisms
alone.~\footnote{For doping concentrations larger than $2\times
  10^{16}$\,cm$^{-3}$ the influence of screening keeps the
  carrier-carrier and carrier-phonon scattering alone from becoming
  more important for the spin dephasing.} For high doping
concentrations beyond $10^{17}$\,cm$^{-3}$, the optically excited
carriers cannot be scattered out of the high $k$ states in which they
are created, because of the high density of relaxed electrons. Then,
by~Eq.~\eqref{Omega-k3}, the precession of the excited electrons
becomes too fast for the timescales of elastic [case(i)] scattering,
inelastic scattering [case(ii)] or their combination [full result], so
that the spin-dephasing time experiences a pronounced
drop.~\cite{zutic:review} The complex behavior of the doping-density
dependence of the spin-dephasing time is thus quantitatively explained
by the interplay of the different elastic and inelastic scattering
mechanisms, and the influence of screening. This picture is
essentially supported by a reanalysis of the data of Fig.~3 in
Ref.~\onlinecite{Shen:chin-phys-lett09}. Last, but not least, we wish
to stress that the numerical results provide a microscopic picture
of motional narrowing in spin dephasing, i.e., an increasing $T_2^*$
with decreasing momentum scattering time, and the breakdown of this
behavior for fast electronic precession. We do not need to resort to
established analytical results for DP spin
dephasing,~\cite{pikus-titkov:opt-orient} which do not work over the
whole density regime and for pronounced non-equilibrium
conditions.~\cite{song:prb02}

\begin{figure}[tb]
\includegraphics[width=0.45\textwidth]{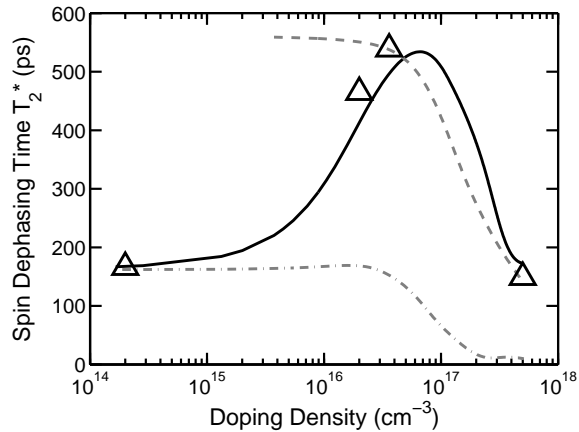}
\caption{Spin dephasing time vs.\ doping density at 4\,K and with 4\,T
  external magnetic field. Shown together with the calculated result
  including all scattering mechanisms (solid line) are curves obtained
  by including \emph{only} impurity scattering (dashed grey line) and
  \emph{without} impurity scattering (dash-dotted grey line). The
  triangles are experimental values. The excited carrier density for
  calculation and experiment is always $N_{\mathrm{exc}}= 2\times
  10^{16}$\,cm$^{-3}$.}
\end{figure}

\section{Conclusion}

We presented results for the spin dynamics of itinerant electrons due
to the DP spin relaxation mechanism for a wide range of doping
concentrations, and under elevated excitation conditions. Our
calculations yield quantitative results for the DP spin relaxation
process, which show the occurrence and breakdown of motional narrowing
in spin dephasing for elevated optical excitation conditions. We find
a maximum of the spin dephasing-time on a sub-nanosecond timescale at
intermediate doping concentrations close to the density where, at low
excitation conditions, a metal-insulator transition occurs. By
numerically investigating the different contributions to the
spin-dephasing time under elevated excitation conditions, we showed
that the maximum of spin dephasing times in the intermediate doping
regime is due to the interplay of elastic and inelastic scattering
mechanisms, as well as screening.

\begin{acknowledgments}
A grant for CPU time from the NIC J\"{u}lich is gratefully
acknowledged. We thank the Graduiertenkolleg 792 of the German Science
Foundation (DFG) for financial support, and Ming-Wei Wu for helpful
discussions.
\end{acknowledgments}


\end{document}